\documentstyle[12pt]{article}

\catcode`\@=11
\long\def\@makefntext#1{
\protect\noindent \hbox to 3.2pt {\hskip-.9pt
$^{{\ninerm\@thefnmark}}$\hfil}#1\hfill}		

\def\@makefnmark{\hbox to 0pt{$^{\@thefnmark}$\hss}}  
\def\ps@myheadings{\let\@mkboth\@gobbletwo
\def\@oddhead{\hbox{}
\rightmark\hfil\ninerm\thepage}
\def\@oddfoot{}\def\@evenhead{\ninerm\thepage\hfil
\leftmark\hbox{}}\def\@evenfoot{}
\def\sectionmark##1{}\def\subsectionmark##1{}}

\renewcommand{\thefootnote}{\fnsymbol{footnote}}

\newcounter{sectionc}\newcounter{subsectionc}\newcounter{subsubsectionc}
\renewcommand{\section}[1] {\vspace*{0.6cm}\addtocounter{sectionc}{1}
\setcounter{subsectionc}{0}\setcounter{subsubsectionc}{0}\noindent
	{\normalsize\bf\thesectionc. #1}\par\vspace*{0.4cm}}
\renewcommand{\subsection}[1] {\vspace*{0.6cm}\addtocounter{subsectionc}{1}
	\setcounter{subsubsectionc}{0}\noindent
	{\normalsize\it\thesectionc.\thesubsectionc. #1}\par\vspace*{0.4cm}}
\renewcommand{\subsubsection}[1]
{\vspace*{0.6cm}\addtocounter{subsubsectionc}{1}
	\noindent {\normalsize\rm\thesectionc.\thesubsectionc.\thesubsubsectionc.
	#1}\par\vspace*{0.4cm}}

\newcounter{appendixc}
\newcounter{subappendixc}[appendixc]
\newcounter{subsubappendixc}[subappendixc]

\renewcommand{\appendix}[1] {\vspace*{0.6cm}
        \refstepcounter{appendixc}
        \setcounter{figure}{0}
        \setcounter{table}{0}
        \setcounter{equation}{0}
        \renewcommand{\thefigure}{\Alph{appendixc}.\arabic{figure}}
        \renewcommand{\thetable}{\Alph{appendixc}.\arabic{table}}
        \renewcommand{\theappendixc}{\Alph{appendixc}}
        \renewcommand{\theequation}{\Alph{appendixc}.\arabic{equation}}
        \noindent{\bf Appendix \theappendixc #1}\par\vspace*{0.4cm}}

\def\abstracts#1{{
\centering{\begin{minipage}{12.2truecm}\vspace*{.1cm}
        \footnotesize\baselineskip=12pt\noindent
	\parindent=0pt #1
	\end{minipage}}\par}}

\renewenvironment{thebibliography}[1]
	{\begin{list}{\arabic{enumi}.}
	{\usecounter{enumi}\setlength{\parsep}{0pt}
\setlength{\leftmargin 1.25cm}{\rightmargin 0pt}
	 \setlength{\itemsep}{0pt} \settowidth
	{\labelwidth}{#1.}\sloppy}}{\end{list}}

\newcounter{itemlistc}
\newcounter{romanlistc}
\newcounter{alphlistc}
\newcounter{arabiclistc}

\newcommand{\fcaption}[1]{
        \refstepcounter{figure}
        \setbox\@tempboxa = \hbox{\footnotesize Fig.~\thefigure. #1}
        \ifdim \wd\@tempboxa > 6in
           {\begin{center}
        \parbox{6in}{\footnotesize\baselineskip=12pt Fig.~\thefigure. #1}
            \end{center}}
        \else
             {\begin{center}
             {\footnotesize Fig.~\thefigure. #1}
              \end{center}}
        \fi}

\newcommand{\tcaption}[1]{
        \refstepcounter{table}
        \setbox\@tempboxa = \hbox{\footnotesize Table~\thetable. #1}
        \ifdim \wd\@tempboxa > 6in
           {\begin{center}
        \parbox{6in}{\footnotesize\baselineskip=12pt Table~\thetable. #1}
            \end{center}}
        \else
             {\begin{center}
             {\footnotesize Table~\thetable. #1}
              \end{center}}
        \fi}

\def\@citex[#1]#2{\if@filesw\immediate\write\@auxout
	{\string\citation{#2}}\fi
\def\@citea{}\@cite{\@for\@citeb:=#2\do
	{\@citea\def\@citea{,}\@ifundefined
	{b@\@citeb}{{\bf ?}\@warning
	{Citation `\@citeb' on page \thepage \space undefined}}
	{\csname b@\@citeb\endcsname}}}{#1}}

\newif\if@cghi

 1
 1
 1

\font\ninerm=cmr9

\def\begi{\begin{itemize}}
\def\endi{\end{itemize}}
\def\beq{\[}
\def\eeq{\]}
\def\begsm{\begin{small}}
\def\endsm{\end{small}}

\begin{document}

\centerline{\normalsize\bf THE DOUBLET-TRIPLET SPLITTING PROBLEM}
\baselineskip=20pt
\centerline{\normalsize\bf AND HIGGSES AS PSEUDOGOLDSTONE BOSONS%
\footnotemark[1]}

\footnotetext[1]
{\baselineskip=14pt
This work is supported in part by funds provided by
the U.S.~Department of Energy (D.O.E.)
under contract \#DE-FC02-94ER40818.
Talk presented at SUSY'95 conference and at the
Joint Meeting of the Johns Hopkins Workshop and PASCOS.
MIT-CTP-2460.  hep-ph/9508208}

\vspace*{0.2cm}

\centerline{\footnotesize LISA RANDALL$^1$}
\baselineskip=12pt
\centerline{\footnotesize and}
\baselineskip=12pt
\centerline{\footnotesize CSABA CS\'AKI$^2$}
\baselineskip=12pt
\centerline{~~~}
\centerline{\footnotesize\it Center for Theoretical Physics,}
\centerline{\footnotesize\it Laboratory for Nuclear Science}
\centerline{\footnotesize\it and Department of Physics,}
\centerline{\footnotesize\it Massachusetts Institute of Technology,}
\centerline{\footnotesize\it Cambridge, MA ~02139--4307}
\centerline{\footnotesize 1: E-mail: lisa@ctptop.mit.edu}
\centerline{\footnotesize 2: E-mail: csaki@mitlns.mit.edu}

\vspace*{0.7cm}
\abstracts{
The doublet-triplet splitting problem is probably the most significant
challenge to supersymmetric GUT theories.  In this talk, we review
potential solutions and their problematic aspects.  We also present a
complete consistent realization of our preferred solution, higgses as
pseudogoldstone bosons, and discuss some distinctive aspects of its
phenomenology.
}

\vspace*{0.7cm}

\normalsize\baselineskip=15pt
\setcounter{footnote}{0}
\renewcommand{\thefootnote}{\alph{footnote}}


Weak scale supersymmetry might be the solution
to the hierarchy problem. Recently in light of
the very accurate unification of gauge couplings,
supersymmetric GUT theories have received a good
deal of attention.
 However, the status of supersymmetric
GUT theories is perhaps not so rosy as it appears.
The essential problem is to find a consistent
picture of the Higgs sector which
requires very light (weak scale) doublets
but very heavy (GUT scale) triplets.
The purpose of this talk is to emphasize the importance
of solving the doublet-triplet splitting problem in establishing
the credibility of supersymmetric GUT theories, and to suggest
a possible solution.

The outline  is as follows. We first review the
basic problem and the status of solutions. We  will see
that most solutions have some problematic
feature, and in general  are quite complicated.  We
then present our favored solution, namely Higgses as pseudo-Goldstone bosons.
 We show that we can build a complete consistent model
of a supersymmetric grand unified theory in which the doublet Higgses
are light   and the triplet  Higgses are heavy.
   We implement a potential which protects
the Higgs mass through  discrete and gauge
symmetries of the potential.  The Higgs mass is protected by an accidental
global symmetry which is accurate at  sufficiently high order in Planck
mass suppressed terms to keep the Higgs boson as light as required
on phenomenological grounds.

The existence of a complete consistent theory is very
important, both from the vantage point of the viability
of SUSY GUTs, and as we will discuss, from the
viewpoint of phenomenology. We will show
that the predictions of our specific model differ
significantly from what would be predicted for a ``generic"
GUT theory, particularly with respect to flavor changing
neutral currents.

\section{Introduction: Doublet-Triplet Splitting, \hfil\break
\phantom{\qquad} The Quagmire of Supersymmetric GUTs}

The major phenomenological motivation for weak scale supersymmetry is
the resolution of the hierarchy problem.  While the Higgs doublet mass
parameters should be in the 100 GeV range, higgsino mediated proton
decay argues (in the simplest models) that the Higgs triplet mass
parameters are in the $10^{16}$ GeV range.  In the minimal SU(5) model,
a parameter must be chosen with accuracy of order $10^{-13}$. Although
this might be technically natural, it is not very compelling as a
solution to the hierarchy problem.

In the absence of simple solutions, we are faced
with the question of whether it makes sense to believe
a model with a  parameter as small as $10^{-13}$.
Clearly, a better solution is warranted and it is important
to explicitly realize it.
For those of you who are still not convinced this is
a problem,  consider the analogous situation for  technicolor.
Technicolor is
{\it not} ruled out by precision electroweak measurements.%
\footnote{For example, the models of
Refs.~\cite{lisa,howard} contain only a minimal technicolor
sector and are therefore acceptable.}
{}~It is the difficulty and complication of making a model
which can generate masses without excessive flavor changing
neutral currents
which  puts technicolor models in disfavor.

Now consider the status of supersymmetric GUTs.
In some ways, the situation is not so different.  There
are serious fundamental problems from the standpoint
of model building. As I have emphasized, the most
serious problem is doublet-triplet splitting, but there
are other aspects one would like to see simply addressed
in a model, such as generating mass, suppressing dangerous
flavor changing neutral currents,  and suppressing dangerous CP violation.
Without a model, it is difficult to believe in the existence
of the theory. Without an economical solution to the
flavor problem in technicolor theories, they were put in
question.
But at least the ``Higgs" sector works. In supersymmetric GUTs,
the
theory is stymied at an even earlier stage.
What should be emphasized here is that the doublet-triplet
splitting problem is not a peripheral issue, but an
essential part of the supersymmetric GUT theory.
Furthermore, any aspect of the phenomenology which
probes the GUT structure   can depend
in an essential way on this part of the theory.  Once
we explicitly construct a model, we will exploit it
to test the sensitivity of certain predictions, and
compare to what might have been derived from a ``generic" GUT.

\section{Review of Proposed Solutions}

There have been a number of solutions proposed to avoid
fine tuning and naturally distinguish the doublet and
triplet Higgs masses.  They generally have clever group
theory structure, which  prevents a doublet
Higgs mass in the renormalizable Lagrangian. However,
as we will see, most models suffer from some unsatisfactory
feature. I will review a few of the more promising suggestions here
and  evaluate their satisfactory and unsatisfactory features.

Before proceeding, let us present some criteria
for a good model. First, there should be
   no unduly small numbers,
that is no tuning of parameters.  This is
a somewhat loosely defined thing--one person's
tuned parameter might be perfectly acceptable
to someone else. We suggest the following subjective,
but probably adequate, definition. Any number which
you would write only in scientific notation
is not permitted.  The number 0.01 might be OK according
to this definition,
but $10^{-5}$    is  not.

It should be emphasized that the
ratio required between the doublet and triplet Higgs
masses is roughly $(M_G/M_P)^4$. So
if one believes that the allowed Planck
suppressed operators are present, it is not sufficient
  to suppress the Higgs mass in the renormalizable
Lagrangian. It must be true that the potential respects
whatever symmetry is necessary to fourth order in $M_P$
suppression.

   A third restriction is that
the model should permit
successful gauge coupling unification
within errors,
both experimental and theoretical.
This is important since the major  phenomenological motivation for GUTs
is the unification of couplings. This is very constraining--
it does not allow for
 even one additional  pair of doublets or triplets.

 Now simply stated, GUT theories
mandate simple relations between the fields in a single
GUT representation. It is very difficult to see how fields
from a single representation can have such different masses.
Not suprisingly the ``solutions" are often very complicated
and are not often entirely adequate, though many are
quite clever. It is very difficult to find a model
in which the doublet-triplet splitting is solved naturally
and in which there is no fine tuning.

In Minimal SU(5), the  best proposed solution is the Missing
Partner Mechanism \cite{missingpartner,grin}.  The idea here
is  to give the triplet a mass through a Dirac
mass term involving a more complicated representation
of SU(5) which has the property that it contains a triplet
but not a doublet.  The 50 is the smallest representation
with this feature.  One therefore constructs the mass terms
\beq
W\supset \lambda 5_H \bar{50}_H \langle 75_H \rangle+\lambda'\bar{5}_H 50_H
\langle {75}_H \rangle
\eeq
so that the triplets, but not the  doublets  are massive.
A model with additional
symmetry to forbid a direct mass term for the $5$ and $\bar{5}$
 incorporates an additional $75$ \cite{grin}.

This is a very nice idea, but seems unlikely to be the
resolution of the dilemma.  There are several problems
with this model.   First of all, the large rank of the representations
is disturbing.  From a theoretical perspective, one
has yet to find string theory examples containing these
large rank representations.  Another problem is that
the gauge coupling grows very rapidly, so that the
theory is strongly coupled not far above the GUT scale.
Although this might be acceptable, it is certainly
a problem at the level of nonrenormalizable operators
which we discuss shortly.

A further problem is that one cannot leave the states
in the remainder of the 50 massless, since they contribute
like an extra doublet pair to unification, which we
know is too much. One can solve this problem for
example by adding a mass term
 $M 50 \bar{50}$ (though this is forbidden
by the symmetry of Ref.~\cite{grin}).  But then there is nothing in
the symmetry structure of the theory which
could forbid the term  $(5) (\bar{5}) (75) ({75})/M_p$
which is the product of two allowed terms in the
superpotential divided by a third and is therefore allowed
by the symmetry, no matter what it is.  If one
believes Planck suppressed operators consistent with
the symmetries are present, the doublet has much
too big a mass.  This problem is exacerbated in
the case the coupling blows up at a low scale, because
it is probably the associated strong scale which would suppress such operators.

A more compact implementation of the Missing Partner
Mechanism was proposed for Flipped SU(5) \cite{ellis}.
The idea is again to pair up the Higgs with ``something else".
Here the something else is a $10_H$ for the $5_H$ and a
$\bar{10}_H$ for the $\bar{5}_H$, the subscript refers to the Higgs sector
to distinguish these fields from the ordinary matter fields.
 These $10_H$ and $\bar{10}_H$ fields are
not dangerous because the nonsinglet nontriplet
fields are eaten when SU(5) breaks. Hence one has
eliminated the necessity for the additional mass term.

The $10_H$ contains a $\bar{3}$ but no color singlet weak doublet.
The $10_H$ and $\bar{10}_H$ get VEVs breaking the SU(5)$\otimes$U(1)
gauge group to the standard
model. The
triplet Higgs in $5_H$ pairs with the triplet in
$10_H$ and the remaining fields
in the  $10_H$  are eaten by the massive vector bosons.

This model might work.   However,
flipped SU(5) is not really a unified model since the gauge group is
SU(5)$\otimes$U(1) which is not a semisimple group.  If
it is embedded in a larger gauge group, the problem should
be solved in the context of the larger gauge group.  The
other feature we find disturbing is that there are many assumptions
about   the vacuum structure.  At tree level,
there is a D-flat, F-flat direction, and the loop corrections have
to be such as to generate the desired minimum.
In Ref.~\cite{ellis}, the correct ratio of gauge and
Yukawa couplings was assumed so that the   VEV's for $10_H$ and $\bar{10}_H$
were at the GUT scale
 while the    $5_H$ and $\bar{5}_H$ VEV's  are small
and the VEVs of an SU(5)  singlet    generated the
  $``\mu"$ term.  It is certainly easier
to  evaluate the vacuum when it is determined at  tree
level, as it will be in our preferred model.

Other solutions have been proposed for models which incorporate
SU(5) as a subgroup, for example SO(10).  The ``missing VEV"
or Dimopoulos-Wilczek mechanism \cite{dw} is probably
the most popular SO(10) solution.
The idea is again to pair the triplet and not the doublet Higgs
with something else so that the triplet, but not the doublet
is massive.
In this model, the way this is done is that
the  VEV  aligns so that the triplet, but
not the doublet, gets a mass.
  (This would not have been possible
in minimal SU(5) due to the tracelessness of the adjoint.)

Again, this mechanism seems very nice at first glance,
but worrisome at second.  For if
    this were all there was, you would have four light doublets,
not two. You need to give the extra doublets
a mass, and the problem is how to do this without
reintroducing a problem with proton decay. A series
of papers by Babu, Barr and Mohapatra \cite{bb1,bb2,bb3,bm}
showed possible ways to make the DW mechanism
into a more complete model.

 The first example \cite{bb1}  had
two  sectors giving VEVs
aligning in different orientations, one  responsible for the triplet  mass,
and one  responsible for the doublet mass.  They
thereby achieved strong suppression of proton
decay. There was an additional field to
complete the breaking of SO(10) to the standard model,
and an additional adjoint to couple the two sectors
together (elimininating a massless Goldstone) without
misaligning the DW mechanism.
 The total field content  in this model is uncomfortably large --
$3(16)+3(10)+3(45)+2(54)+\bar{16}+16$, leading to  fairly big
threshold corrections and  the blowing up of
the gauge  coupling before $M_{Pl}$.  Other problems
with this particular model was that some operators
which would have been allowed by the symmetries of the model
needed to be forbidden, and that nonrenormalizable
Planck mass suppressed operators could be dangerous.

This last problem was  addressed in their second model,
 where they sacrifice strong suppression of proton
decay   but generate a natural model,
in the sense that they include
all operators permitted by their assumed
symmetry structure.  The field content of this model was
$3(16)+2(10)+3(45)+(54)+(\bar{126})+(126)$.  Discrete symmetries
were sufficient to forbid any unwanted terms from the potential.
However, the field content was still quite large, and high
rank representations were required.

The third model incorporated a smaller field content and
no high rank representations, so it
  should be more readily obtainable from string models.
In this model, the authors achieved the  DW form with higher dimension
operators, so no (54) was required.
There was a  $\bar{16}+16$ to complete the  breaking to the standard model.

However, without three adjoints, there were intermediate scale pseudo-Goldstone
bosons. The authors resolved this problem by cancelling
the fairly large corrections to unification of couplings (due
to the light charged fields) by large threshold corrections.  Although
this might work, it is at the edge of parameter space.

Another nice model based on SO(10) is the model of Babu and
Mohapatra \cite{bm} which allows for a 10-16 mixing and
therefore a  Higgs sector  which distinguishes the up and down
quark masses.
    However, this
model had a few small (but not very small) parameters, a flat
direction and therefore vacuum degeneracy at tree level, extra
singlets, and a complicated superpotential.

To summarize, there are some interesting models in the literature,
primarily based on clever group theory structure. However
most models suffer from one of the following problems.
\begi
\item
   There is the problem of actually implementing the potential to get
the desired minimum and
light Higgses.
 The minimum can sometimes be   destabilized with higher order terms.
  Also some models have flat directions so the vacuum
needs to be carefully thought through.
\item It is necessary to ensure the  light particle
spectrum is compatible with gauge coupling unification.
  Most solutions rely on pairing up the  triplet higgsinos (not
doublet) with ``something else".
  ``Something else" can be a problem
(with gauge unification).

\item The particle  representation is  cumbersome. This leads
to the questions of whether it is derivable from strings or
whether the coupling blows up before the Planck scale.
In any case, models with large particle content seem
unappealing and unlikely.
 \endi

The problem is clear.
Minimal SU(5)  relates doublets and triplets!  Almost always, the solution
relies on a compromise at  the edge of parameter space
or tuned parameters or setting some couplings
to zero in the potential.   This
is a good  introduction to the Higgses as pseudo-Goldstone bosons model
which we will argue is an exception to the discussion above.
Rather than relying on pairing the Higgs in complicated
ways, the theory relies on a spontaneously broken
symmetry under which the Higgses are Goldstone bosons.
This distinguishes the doublets from the triplets in a very
nontrivial way, so that it is natural to
obtain light doublets when the remaining fields
are heavy.  The originally proposed model \cite{inoue,Ans,Bar1}
involved gauged
SU(5) symmetry and a  global SU(6) symmetry
which was implemented by tuning potential parameters.
A better model \cite{Zur,Bar2,Bar3,Bar4} was later proposed which
admits the possibility for justifying the large global
symmetry with discrete symmetries.
In fact, as we will see, one can construct
a simple model to implement this idea \cite{bcr}.

The general idea behind the Higgs as Goldstone scheme
is to implement a global symmetry on the Higgs sector
which is broken (explicitly) by the Yukawa couplings to
matter.  The masses of the pseudo-Goldstone bosons are protected from large
loop corrections by the nonrenormalization theorem.
The Higgs sector can be distinguished by
matter parity and consists of the adjoint
 ($\Sigma$), the fundamental representation
  ($H$)  and the
antifundamental ($\bar{H}$), where this refers to their
representation under the gauge symmetry. Then additional
{\it global} symmetry is ensured
by assuming the superpotential is of the form
\beq
W(\Sigma, H, \bar{H})=W(\Sigma)+W(H,\bar{H})
\eeq
At the renormalizable level, this can be achieved by
forbidding the  coupling $\bar{H} \Sigma H$. At higher
orders, many other couplings must be  forbidden.

 By far the nicest implementation of this
idea (here we say how we would like
the structure to work without explicitly implementing
the potential which we do later) is  based
on extending minimal SU(5) to the gauge group SU(6),
with SU(6) $\otimes$ SU(6)
global symmetry. We will refer to this as the SU(6) model.
The Higgses are then an adjoint, $\Sigma$ which is a $35=24+ 6+\bar{6}+1$,
a fundamental $H$, in a $6=5+1$,
and an $\bar{H}$, in a $\bar{6}=\bar{5}+1$ where we have given
the SU(5) decompositions.
The accidental symmetry is realized if mixing terms of the form
$\bar{H}\Sigma H$ are not present in the superpotential. If the fields
$\Sigma$ and $H,\bar{H}$ develop VEV's of the form
\begin{equation}
\label{svev}
\langle \Sigma \rangle = V \left( \begin{array}{cccccc} 1 & & & & & \\
& 1 & & & & \\ & & 1 & & & \\ & & & 1 & & \\ & & & & -2 & \\
& & & & & -2 \end{array} \right),
\end{equation}
\begin{equation}
\label{hvev}
\langle H \rangle =\langle \bar{H} \rangle = U \left( \begin{array}{c}
1\\0\\0\\0\\0\\0 \end{array} \right),
\end{equation}
then one of the global SU(6) factors breaks to
 SU(4) $\otimes$ SU(2) $\otimes$ U(1),
while the other to SU(5). Together, the VEV's break the gauge group to
SU(3) $\otimes$ SU(2) $\otimes$ U(1).

The Goldstone bosons (GB's) coming from the breaking
SU(6) $\rightarrow$ SU(4) $\otimes$ SU(2) $\otimes$ U(1)
are (according to their SU(3) $\otimes$ SU(2) $\otimes$ U(1)
transformation properties):
\begin{equation}
(\bar{3},2)_{\frac{5}{6}}+(3,2)_{-\frac{5}{6}}+(1,2)_{\frac{1}{2}}+
(1,2)_{-\frac{1}{2}},
\end{equation}
while from the breaking SU(6)$\rightarrow $SU(5) the GB's are
\begin{equation}
(3,1)_{-\frac{1}{3}}+(\bar{3},1)_{\frac{1}{3}}+(1,2)_{\frac{1}{2}}
+(1,2)_{-\frac{1}{2}}+(1,1)_0.
\end{equation}
But the following GB's are eaten by the heavy vector bosons due to the
supersymmetric Higgs mechanism (the gauge symmetry is broken from SU(6) to
SU(3) $\otimes$ SU(2) $\otimes$ U(1)):
\begin{equation}
(3,1)_{-\frac{1}{3}}+(\bar{3},1)_{\frac{1}{3}}+(3,2)_{-\frac{5}{6}}
+(\bar{3},2)_{\frac{5}{6}}+(1,2)_{\frac{1}{2}}+
(1,2)_{-\frac{1}{2}}+(1,1)_0.
\end{equation}
Thus exactly one pair of doublets remains uneaten which can be identified
with the Higgs fields of the MSSM. One can show that the uneaten doublets
are in the following combinations of the fields $\Sigma ,H,\bar{H}$:

\begin{equation}
\label{higgs}
h_1=\frac{U h_{\Sigma}-3V h_H}{\sqrt{
9V^2+U^2}},
\end{equation}
\begin{equation}
\label{higgs2}
h_2=\frac{U \bar{h}_{\Sigma}-3V
\bar{h}_{\bar{H}}}{\sqrt{
9V^2+U^2}},
\end{equation}
where  $h_H$ and $\bar{h}_{\bar{H}}$ denote
the two doublets living in the SU(6) field $H$ and $\bar{H}$, while
$h_{\Sigma}$ and $\bar{h}_{\Sigma}$ denote
the two doublets living in the SU(6)
adjoint $\Sigma$.
In order to  maintain the correct prediction for
$\sin^2\theta$, we need to have
$\langle \Sigma \rangle \sim M_{GUT}$, $\langle H \rangle = \langle \bar{H}
\rangle > \langle \Sigma \rangle$.

It is important to note that the
triplets in   $H$ and $\bar{H}$ are  eaten and not dangerous.
Only the $\Sigma$ triplet needs to be made  heavy
in the superpotential to avoid proton decay.

Having established the desired vacuum and symmetry structure,
we now have to face the question of whether
such a model exists.  In fact, we would
once again like to see a model with the more
stingent requirement that the Higgs is massless
up to order $(M_G/M_{Pl})^4$.  Furthermore
we will take the point of view that only gauge and discrete
symmetries are exact. To obtain the desired structure   therefore requires
an accidental symmetry which is respected in
nonrenormalizable terms to fourth order  in $1/M_{Pl}$.
We also
  require that the  potential has no flat directions so that
the minimum is
determined and is the desired one.  In order
to show that it is a viable model we also  require that the model
  can be extended to fermion masses.
We emphasize that the existence of such a model
 is important because there are no other complete
consistent  and simple models. The model   \cite{bcr} we
constructed has simple field content, no small
parameters, and  the above properties.

 Before actually presenting an example of a model,
let us first understand  the difficulty.
 Suppose the potential is
such that the minima of $\Sigma$ and $H$ are determined.
We then require at least four terms in the potential (one can not have
nonzero VEVs for  $\Sigma$ and $H,\bar{H}$ with just three terms).

\beq
\frac{1}{M_{Pl}^{a-3}} {\rm Tr} \Sigma^a + \frac{1}{M_{Pl}^{b-3}}
 {\rm Tr} \Sigma^b,\label{bal}
\eeq
\beq
\frac{1}{M_{Pl}^{2c-3}}(\bar{H}H)^c+\frac{1}{M_{Pl}^{2d-3}} (\bar{H}H)^d
\eeq
But then the symmetries allow
\beq
\frac{1}{M_{Pl}^{2c+b-a-3}} (\bar{H}\Sigma^{b-a}H) (\bar{H}H)^{c-1}
\eeq
 Notice that the dangerous
term involving both $\Sigma$ and $H$ is suppressed by   precisely the
ratio of two superpotential terms. But
at the minimum, all the terms in the potential
should be of the same order of magnitude.  This
means that the ratio of fields is of the same
order of magnitude as the ratio of the coefficients
of the operators. To get the required suppression
of the Higgs mass would then require a ratio
of coefficients of order $(M_W/M_{Pl})$! This is
precisely what we are trying to avoid--that
is tuning the Higgs mass to be small.

So we need to explore the possible loopholes in the
above reasoning.   One possibility is to add more fields,
so that one cannot make a holomorphic function with
positive powers of the fields which is allowed by
the existing symmetries. But to eliminate the
flat directions then requires more terms in the superpotential.
 Furthermore
we do not want very high dimension operators
in the $\Sigma$ potential (because
when the VEV is of order $M_G$, the triplets
in the $\Sigma$ will get too small a mass,
suppressed by $(M_G/M_{Pl})$ to a large power).
Also, the dimensions of the operators in the $\Sigma$
or $H$, $\bar{H}$ potential should not
be very different, or else the ratio of couplings
will be large (or small) so that all terms
in the potential can be of the same order
of magnitude at the minimum.    But with sufficiently many fields and
low dimension terms in the potential
to eliminate flat directions, one can generally make invariant operators
by more complicated extensions of the above argument.   We
found no examples where we obtained a satisfactory
minimum at tree level and all dangerous operators were forbidden.

We conclude we need  a small number to make a model.
Fortunately we know there exists a small number, namely
the ratio of the weak scale to the Planck scale. Even
without knowing how this ratio arises, we know this
small number is present in any satisfactory model.

The alternative for a small number is $0$. One can
also construct models which incorporate fields with zero expectation
value, so that the dangerous terms involving the $\Sigma$ and $H$
fields vanish.

Several models involving one or the other of these ideas were
presented in Ref.~\cite{bcr}.  We present the simplest
of the examples in the following section.

\section{A Model}

The field content is the minimal field content
required, namely $H$, $\bar{H}$, and $\Sigma$.  We
impose an additional discrete symmetry
  $\bar{H}H$: $\bar{H} H\to e^{2\pi i/n} \bar{H} H$.

The superpotential is
\beq
\label{softpot}
W=\frac{1}{2}M{\rm Tr} \Sigma^2+\frac{1}{3}\lambda{\rm Tr}
 \Sigma^3 +\alpha\frac{(\bar{H}H)^n}{M_{Pl}^{2n-3}}
\eeq
and the potential after the inclusion of the soft breaking terms is given by
 \begin{eqnarray}
\label{softbr}
& & V(\Sigma ,\bar{H},H)={\rm Tr}|M\Sigma +\lambda\Sigma^2-\frac{1}{6}\lambda
{\rm Tr}\Sigma^2|^2 \nonumber \\
&&+\frac{n^2\alpha^2}{M_{Pl}^{4n-6}} (\bar{H}H)^{2n-2} (|H|^2+|\bar{H}|^2)+
\nonumber
\\ & & A m \lambda {\rm Tr}\Sigma^3+ A'm \alpha
\frac{(\bar{H}H)^n}{M_{Pl}^{2n-3}}+ BMm\Sigma^2\nonumber \\
&&+m^2({\rm Tr}\Sigma^2 +|H|^2+|\bar{H}|^2) +\;
{\rm D-terms} \nonumber \\
\end{eqnarray}
Notice that at the minimum for $H$, a high dimension operator
is balanced against the  soft supersymmetry breaking terms.
 There is a minimum with
\beq
\langle H\rangle=\langle \bar{H} \rangle=\left(m \over M_{Pl}\right)^{1
\over 2n-2} M_{Pl}
\eeq
  For $n=4,5,6$, we get $\langle H \rangle \approx 1.5 \cdot 10^{16}, 6\cdot
10^{16}, 2 \cdot 10^{17}$ GeV.  The first
 mixing term allowed by $Z_n$ is $\frac{1}{M_{Pl}^{2n-2}} (\bar{H}H)^{n-1}
(\bar{H}\Sigma H)$.
This gives mass to the pseudo-Goldstone bosons which is less than the
weak scale, and therefore safe.
 Notice it was essential that $H$ mass term was small for naturalness
of the model. This was only permitted because the triplets
in $H$ and $\bar{H}$ were eaten.

We emphasize that this model had no   fields
other than an adjoint, fundamental,
and antifundamental, which we expect
to be present in any SU(n) model which
breaks to the standard model and gives
the necessary fermion masses.   However,
to accomodate fermion masses in our context
and to generate a successful mass texture, we
found it useful to incorporate an additional adjoint field.
  With  an additonal  discrete $Z_3$ symmetry (under which
$\Sigma_1 \rightarrow e^{2\pi i/3}\Sigma_1$, $\Sigma_2
\rightarrow e^{-2\pi i/3}\Sigma_2$, $\bar{H}H\rightarrow \bar{H}H$),
the superpotential
takes the form
 \beq
\label{goodsup}
M{\rm Tr}\Sigma_1\Sigma_2 +\frac{1}{3}\lambda_1{\rm Tr}\Sigma_1^3
+\frac{1}{3} \lambda_2  {\rm Tr}\Sigma_2^3 +\frac{\alpha}{M_{Pl}^{2n-3}}
(\bar{H}H)^n
\eeq
This is useful for constructing a mass model in which the masses
and mixing angles are naturally of order of magnitude of the
ratio of VEVs and the Planck scale \cite{bcr}. These are
enforced through discrete symmetries acting on the
fermions.  One interesting aspect of these models
is that the top quark is in a distinct representation
from the other up type quarks \cite{Bar3,bcr}. This is required
in order to give a renormalizable Yukawa coupling
to the top quark so it can get a sufficiently
large mass.  Another interesting aspect
of the mass models is that $b-\tau$ unification
occurs naturally, because there is a unique
operator which gives a mass to the $b$ and $\tau$
(so there are no Clebsches distinguishing the mass).

\section{Phenomenology}

Because of the enlarged gauge group and the restricted
Higgs sector, the phenomenology of the SU(6) model
operates very differently from the minimal SU(5) model.
However, since there is probably not a minimal SU(5) model,
since any GUT  model which solves the doublet-triplet
splitting problem is likely to have additional structure, it is
important to explore the phenomenology of a model with a satisfactory
Higgs sector.  These predictions can differ significantly
from ``generic" results.

Here we focus on the violations of lepton
flavor in supersymmetric unified theories \cite{bhs}.  Barbieri,
Hall, and Strumia
point out that in supersymmetric GUTs there is large
flavor violation in the {\it lepton} sector due to $\lambda_t$
and that it is communicated to physics at low energies through
the scalar partners.  In principle, one can test SUSY GUTs
through flavor changing processes, such as
  $\mu \to e \gamma$. They worked out the predictions
for the minimal GUT theories. However,
the SU(6) model predictions   look very different as a function
of the couplings and SUSY soft parameters from
the minimal SU(5) case.
Let us focus for the moment on what distinguishes
the predictions of the SU(6) model from  those of a ``generic"
GUT.  First of all, the top quark is in a different gauge
representation from the other up type quarks, so
in principle there are additional
  flavor changing effects from {\it gauge} interactions.
It turns out however these are smaller than those
due to the top Yukawa coupling. Second, there
are
 different possible contractions of higher dimension
operators so there are   different mixing angles
for the  leptons and down type quarks, reducing
the predictability of the flavor changing lepton process
(related to unknown mixing angles).
Another distinguishing aspect of our model is
that there are
 additional potential flavor dependent Yukawa
couplings from interactions with heavy fermions (which are necessarily
present in the theory for the reason of anomaly cancellation).
 These can however be naturally suppressed
by discrete symmetries \cite{cr}.

What turns out to be the most important numerical
difference to the prediction is the larger gauge
group requiring larger representations.
This makes everything run much faster,
since essentially the counting factors
on loop diagrams are bigger.

Let us now consider the predictions.
 First we look
at the top Yukawa coupling, $\lambda_t$.  There is
an upper bound in this model from two things--first
we require the top Yukawa to be perturbative up to the
Planck scale, which gives a bound $\lambda_t<1.12 $
(compared to $\lambda_t<1.56  $ in minimal SU(5).
 Second, we require that the stau mass
remains positive. In minimal SU(5) the running
is slower, so there is usually no such a constraint.
 There is also a lower bound on $\lambda_t$. This
comes from requiring that $b$-$\tau$ unification
works. In our model, it is easy to
make $b$-$\tau$ unification work to a level of a few percent,
but difficult to make it work with greater than 10\% error
without completely destroying the prediction.  With
only this much leeway, $\lambda_t$ cannot be too small,
  greater than about 0.9.

Now with $\lambda_t$ constrained to be so small, minimal
SU(5) would predict unobservably small flavor violations.
However, because in SU(6) the masses run so much more
quickly, we find we still predict potentially testable flavor
changing lepton processes over most of parameter space. The details of this
analysis will be given in Ref.~\cite{cr}.  We find
for comparable $\lambda_t$, the rate for $\mu \to e\gamma$
is an order of magnitude
greater than for minimal SU(5).

\section{Conclusions}

Supersymmetric GUTs seem very nice, but they
are theoretically problematic. The most important
issue from a model building perspective is to
understand how the doublet-triplet splitting problem
can be resolved. Without a solution, the
existence of the theory is difficult to support.

Most solutions have some problem.  However,
the Higgs as pse\-u\-do-\-Gold\-stone boson solution
appears to be an exception.  We have constructed
a rather simple model where the Higgs is light
because of additional accidental global symmetry
present in the theory.  This gives a natural
resolution to the doublet-triplet splitting (and the $\mu$)
problem.  We have a complete consistent example.
Moreover, with an explicit realization of the model,
one can explore its phenomenological consequences.
This gives a new and maybe even more realistic
perspective on  low energy phenemenology. It
is important to understand the range of possibilities
which follow from a complete model.

One might also interpret the difficulty in constructing
GUT models as indicative that   the theory is not unified
below the Planck scale.  In this case, we will ultimately
want to understand the discrepancy between the
unification scale and the Planck scale.  If
we are to apply the same standards applied to the GUT
models, we would also
want to better understand the string vacuum and resolve the
problem of moduli proliferation and  the associated vacuum
degeneracy.
After all, all the solutions appeared beautiful  until
one tried to implement them explicitly in realistic
nonfinetuned natural models which gave the correct
value for $\sin^2 \theta$.

\section {Acknowledgements}
We thank our  collaborator  Zurab Berezhiani.
We would also like to thank  Greg Anderson, Marcela Carena, Diego Casta\~no
and Carlos Wagner for useful conversations.

\end{document}